\documentclass[aps,prl,twocolumn,showpacs,nofootinbib,groupedaddress,superscriptaddress]{revtex4}

\usepackage{amssymb,hyperref}
\usepackage{times,subfigure}

\newcommand{\nc}{\newcommand}
\nc{\ox}{\otimes}

\newcommand{\ii}{{\rm i}}

\usepackage{graphicx}
\usepackage{bm}
\usepackage{amsmath}
\usepackage{amsthm}

\newtheorem{theorem}{Theorem}

\newtheorem{definition}{Definition}

\newcommand{\ket}[1]{|#1\rangle}
\newcommand{\bra}[1]{\langle#1|}
\newcommand{\proj}[1]{\ket{#1}\bra{#1}}
\newcommand{\unidim}[2]{\ket{#1}\bra{#2}}

\newcommand{\ot}[0]{\otimes}
\newcommand{\beq}{\begin{equation}}
\newcommand{\eeq}{\end{equation}}

\newcommand{\Tr}{{\rm Tr}}

\newcommand{\cI}{{\mathcal{U}}}

\nc{\CC}{{{\mathbb C}}}
\nc{\EE}{{{\mathbb E}}}

\begin{document}

\title{The quantumness of correlations revealed in local measurements exceeds entanglement}

\author{Marco Piani}
\affiliation{$\mbox{Institute for Quantum Computing and Department of Physics and Astronomy, University of Waterloo, Waterloo ON
N2L 3G1, Canada}$}

\author{Gerardo Adesso}
\affiliation{$\mbox{School of Mathematical Sciences, The University of Nottingham, University Park, Nottingham NG7 2RD, United Kingdom}$}
\begin{abstract}
  We analyze a family of measures of general quantum correlations for composite systems,  defined in terms of the bipartite entanglement necessarily created between systems and apparatuses during local measurements. For every entanglement monotone $E$, this operational correspondence provides a different measure $Q_E$ of quantum correlations.
Examples of such measures are the relative entropy of quantumness, the quantum deficit, and the negativity of quantumness.
In general, we prove that any so defined quantum correlation measure is always greater than (or equal to) the corresponding entanglement between the subsystems, $Q_E \ge E$, for arbitrary states of composite quantum systems.
We analyze qualitatively and quantitatively the flow of correlations in iterated  measurements, showing  that general quantum correlations and entanglement can never decrease along von Neumann chains, and that genuine multipartite entanglement in the initial state of the observed system always gives rise to genuine multipartite entanglement among all subsystems and all measurement apparatuses at any level in the chain.
Our results provide a comprehensive framework to understand and quantify general quantum correlations in multipartite states.
\end{abstract}


\date{December 15, 2011}

\maketitle

\noindent {\it Introduction.}---
The quantum world differs from our familiar classical world in many, interrelated ways \cite{wehner}. Quantum laws forbid basic tasks such as cloning \cite{nocloning} yet enable certain information processing feats otherwise unfeasible with purely classical resources~\cite{nielsen}.  In particular, quantum correlations differ from classical ones. Such a difference can assume the striking traits of entanglement~\cite{EPR, schroedinger, hororev} and non-locality \cite{bell}, or the subtler features of quantum discord~\cite{discord}. The latter captures a more general signature of non-classicality of correlations (being present also in almost all unentangled states \cite{ferraro}) that stems from the non-commutativity of quantum observables, and can be revealed in local measurements; its characterization and applications have recently attracted much attention~\cite{manydiscord,merali}. Discord or, in general, quantum correlations different from entanglement have been linked to the advantage, over classical scenarios, of quantum algorithms for communication \cite{dattacommun}, information locking \cite{locking}, metrology \cite{discordmetro},  and especially computation in the presence of noise \cite{dattaqcetal}.

\begin{figure}[t]
\centering \includegraphics[width=8.5cm]{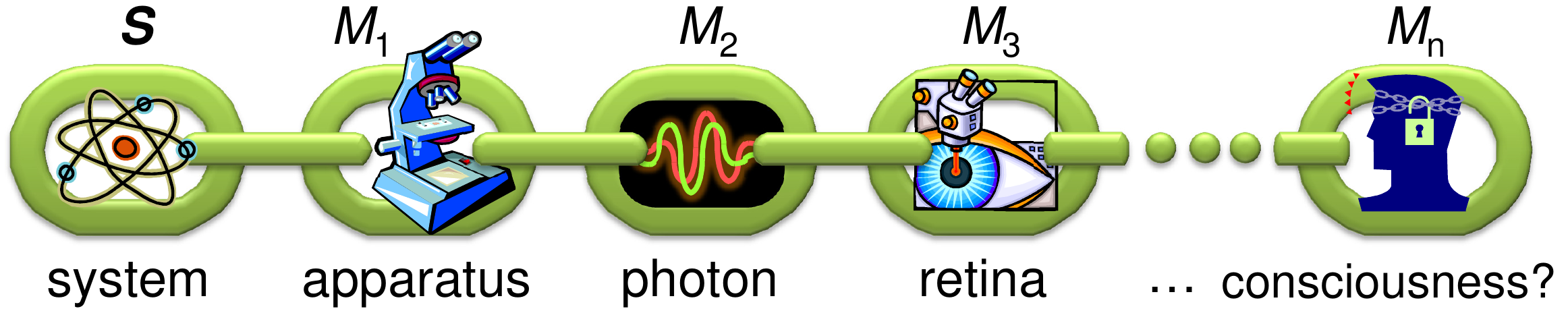}
\caption{(Color online) Graphical depiction of a von Neumann chain.}
\label{fig:vNchain}
\end{figure}

Clarifying the relation between ``classical'' and ``quantum'' is of paramount importance from the practical point of view of information processing, as a better understanding of all genuinely quantum effects can only lead to them being more efficiently exploited. Nonetheless, another more fundamental reason to study the non-classicality of correlations is that we do not fully understand the quantum-to-classical transition \cite{zurekrmp}; it is not clear how and at which scale quantum mechanics actually leads to an everyday world where (macroscopic) objects appear to be ``here'' or ``there'', but not ``here and there'' as instead allowed by the superposition principle. This issue of the emergence of the classical from the quantum is intimately related to the ``measurement problem'' which has puzzled physicists, mathematicians and philosophers since the birth of quantum mechanics a century ago. A potential solution of the measurement problem is in terms of decoherence, that is in terms of correlations established with an ``environment'' on which the ``observer'' has little or no control (for a review on this topic, see~\cite{zurekrmp}). At the same time, it is exactly through the establishment of correlations between the observed quantum system and a measurement apparatus---and finally the observer---that we can describe the measurement to take place. In this respect, the measurement problem becomes that of finding how the ``quantum information'' contained in a quantum system becomes correlated to the ``classical information'' in the mind of the observer.  In analyzing such an issue, von Neumann~\cite{vonneumann} considered what later became known as ``von Neumann chain'': a sequence of interacting physical systems, starting with the quantum system to be measured and ending with the observer (see Fig.~\ref{fig:vNchain}).

In this paper we provide quantitative constraints regulating the establishment of quantum correlations with apparatuses during a measurement process, and their flow along von Neumann chains. Combining and extending the approaches recently put forward in Refs.~\cite{bruss,ouracti}, we quantify the quantum correlations $Q_E$ between subsystems $S_k$ of a composite quantum system $\bf S$, in terms of the minimum entanglement $E$ generated between the whole system ${\bf S}$ and a generally composed measurement apparatus $\bf M$ which is probing a subset of the subsystems of $\bf S$ locally---that is, a local independent measurement on each probed $S_k$ is performed through an interaction with a local apparatus $M_k$ that is part of ${\bf M}$. Here we show, for any entanglement measure $E$, that according to such a mapping the key inequality $Q_E(\rho) \ge E(\rho)$ holds for all quantum states $\rho$. This proves that measurement processes provide a natural and insightful framework for the study of the non-classicality of correlations: for every chosen entanglement monotone $E$, one obtains a different, operational  measure $Q_E$ of general quantum correlations, which incorporates and generally exceeds the entanglement  (measured by $E$) between the subsystems of arbitrary quantum states.  Based on these results, that definitely elucidate the (so far unclear \cite{unclear,discord}) interplay between entanglement and general quantum correlations in composite systems, we further show that quantum correlations and entanglement can never decrease along von Neumann chains, and that all the links of the chain exhibit {\it genuine} multipartite entanglement if and only if the subsystems are genuinely (multipartite) entangled---and not just quantumly correlated---in the initial state of the system $\bf S$.

\smallskip
\noindent {\it Preliminaries.}---
We begin by setting our notation and recalling  a number of definitions. Let ${\bf S}$ be a quantum system partitioned into $n$ (finite-dimensional) subsystems and let $\mathcal{U}=\{1,2,\dots,n\}$; we denote by $S_I=\{S_k|k\in I\}$, with $I\subseteq\mathcal{U}$, a subset of the $n$ subsystems $S_1,S_2,\dots,S_n$ of ${\bf S}\equiv S_{\mathcal{U}}$.
We denote by $\cI\backslash I$ the complement of set $I$ within $\cI$.
With respect to the notion of classicality of correlations, we adopt the following definition~(see also \cite{eric}).
\begin{definition}[Classically correlated states] \label{defCC}
A state $\rho_{S_{\cI}}$ is ``classically correlated (CC) with respect to local measurements on subsystems $S_I$'', or, in other words,  ``classical on subsystems $S_I$'', or,  in short-hand notation, $\rho_{S_{\cI}}$ is ``{\em  $I$-CC}'', if there exists a choice of local complete von Neumann measurements 
on each subsystem $S_k$, $k \in I$ such that  $\rho_{S_{\cI}}$ is left invariant under such measurements.
Equivalently, $\rho_{S_{\cI}}$ is {\em  $I$-CC} if there exists a choice of a local orthonormal basis $\{\ket{i}_{S_k}\}$ for each $S_k$, $k\in I$, such that \begin{eqnarray} \label{clasdefeqn}
\rho_{S_\cI}&=&\sum_{i_{k_1},i_{k_2},\ldots,i_{k_{|I|}}} \proj{i_{k_1}}_{S_{k_1}}\otimes \proj{i_{k_2}}_{S_{k_2}} \\
&\otimes&\cdots \otimes\proj{i_{k_{|I|}}}_{S_{k_{|I|}}}\otimes \rho^{i_{\vec{k}}}_{S_{\cI\backslash I}}\,, \nonumber
\end{eqnarray} with $\rho^{i_{\vec{k}}}_{S_{\cI\backslash I}}$ unnormalized states of the other subsystems $S_{\cI\backslash I}$.
\end{definition}
The above definition provides a finer graining between the conventional categories of strictly classically correlated (i.e., ${\cI}$-CC) and genuinely quantumly correlated (i.e., $\emptyset$-CC) states, including all possible intermediate types of so-called classical-quantum states \cite{pianibroadcast}.

\begin{figure}[t]
\centering \includegraphics[width=7cm]{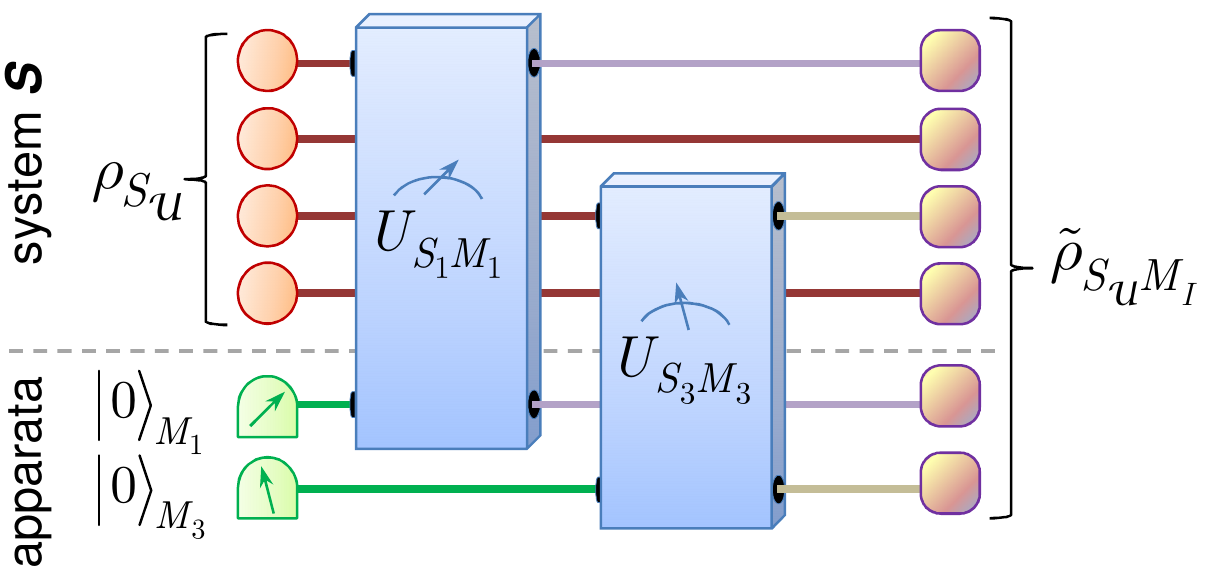}
\caption{(Color online) Construction of the pre-measurement state $\tilde{\rho}_{S_\cI M_I}$ [Eq.~\eqref{statepremeq}], for ${\bf S} \equiv S_\cI = \{S_1, S_2, S_3, S_4\}, I=\{1,3\}, M_I = \{M_1, M_3\}$, and $U_{S_k M_k}$ of the form (\ref{eq:uniinteraction}).}
\label{fig:premeas}
\end{figure}

A complete von Neumann measurement in a basis $\{\ket{i}_{S_k}\}$ can be realized on a subsystem $S_k$ by letting $S_k$ interact with a measurement apparatus $M_k$. Assuming $M_k$ is initially in some fixed but otherwise arbitrary initial pure state $\ket{0}_{M_k}$, and that $S_k$ and $M_k$ interact through a unitary $U_{S_k M_k}$, one finds that the latter must be of the type 
\beq
\label{eq:uniinteraction}
U_{S_k M_k}\ket{i}_{S_k}\ket{0}_{M_k}=\ket{i}_{S_k}\ket{i}_{M_k},
\eeq
up to a local unitary on $M_k$, if we impose that the interaction realizes the projective measurement, that is,
$\Tr_{M_k}(U_{S_k M_k} \rho_{S_k}\otimes  \proj{0}_{M_k} U_{S_k M_k}^\dagger )= \sum_i \proj{i}_{S_k} \rho_{S_k} \proj{i}_{S_k}$,
for all states $\rho_{S_k}$ of $S_k$. The state
\begin{equation}\label{statepremeq}
\tilde{\rho}_{S_k M_k}=U_{S_k M_k} \rho_{S_k}\otimes  \proj{0}_{M_k} U_{S_k M_k}^\dagger\,,
 \end{equation}
 before tracing out the apparatus, is sometimes called the {\em pre-measurement state}~\cite{zurekrmp}. The notion is trivially generalized to the case where local projective measurements are performed on any subset $I$ of $\cI$: in such a case we will refer to a pre-measurement state $\tilde{\rho}_{S_\cI M_I}$ of the whole system plus the set of apparatuses coupled to subsystems $S_I$ (see Fig.~\ref{fig:premeas}). Such a pre-measurement state depends of course on the choice of local bases in which the local measurements take place. By considering the entanglement properties of $\tilde{\rho}_{S_\cI M_I}$ across the $S_\cI:M_I$ bipartition we can formulate the following theorem, straightforwardly generalizing Refs.~\cite{bruss,ouracti}:
\begin{theorem} \label{teo1}
A state $\rho_{S_\cI}$ is {\em $I$-CC} if and only if there exist a set of local complete von Neumann measurements on subsystems $S_I$ such that the pre-measurement state $\tilde{\rho}_{S_\cI M_I}$ is  unentangled across $S_\cI:M_I$.
\end{theorem}
Theorem \ref{teo1} provides a clear operational interpretation of $I$-CC states as the only states which do not necessarily give rise to the creation of entanglement with a set of apparatuses during local pre-measurements of the subsystems $S_I$. When, on the other hand, such entanglement is necessarily created, its minimum amount (measured by any chosen monotone $E$), where the minimization is over the choice of the local measurements, can be regarded as a measure of quantum correlations $Q_E$ in the original state $\rho_{S_\cI}$ of the system \cite{bruss,ouracti}.
We can thus formulate the following quantitative definition for a general family of measures of quantum correlations.
\begin{definition}[Quantumness of correlations] \label{defQ}
The measure $Q_{E}^{(S_I)}$ of quantumness of correlations---or, simply, quantum correlations---among all the subsystems of the system $S_\cI$ in the state $\rho_{S_\cI}$, revealed by local measurements on the (subset of) subsystems $S_I$, and corresponding to the  entanglement measure $E$, is defined by
\beq\label{defqeq}
Q_{E}^{(S_I)}({\rho_{S_{\cI}}}):=\min_{\{\{\ket{i}_{S_k}\}|k\in I\}} E_{S_{\cI}:M_I}({\tilde{\rho}_{S_\cI M_I}}),
\eeq
where the minimum runs over all choices of local bases (equivalently, local complete von Neumann measurements) for subsystems $S_I$, and the bipartite entanglement measure $E$ is calculated between the systems $S_{\cI}$ and $M_I$ in the pre-measurement state $\tilde{\rho}_{S_\cI M_I}$ [Eq.~\eqref{statepremeq}], which depends implicitly on the local bases choice.
\end{definition}
According to Definition~\ref{defQ}, for a bipartite system ${\bf S} \equiv S_{AB}$ ($n=2, S_1\equiv A, S_2 \equiv B$), if $E$ is chosen to be, e.g., the distillable entanglement $E_D$ \cite{distillable,hororev}, then $Q_{E_D}^{(AB)}$ amounts to the (two-way) relative entropy of quantumness \cite{ouracti,req} and $Q_{E_D}^{(A)}$ amounts to the one-way information deficit \cite{bruss,deficit}. Other instances of measures falling in the category of Definition \ref{defQ} are, e.g., the so-called negativity of quantumness \cite{ouracti} (also known as minimum entanglement potential \cite{melo}) which corresponds to picking $E$ to be the negativity ${\cal N}$ \cite{hororev}, and the geometric measure of quantumness \cite{brussnoise}. All the mentioned measures, among all, have been independently proposed in different contexts (ranging from thermodynamics to geometric approaches) and have well understood definitions that capture various, intertwined features of non-classical correlations in composite systems. Our framework, which builds on and generalizes the ideas of \cite{bruss,ouracti}, provides universal and physically justified {\em operational} interpretations to those and a whole lot of infinitely-many possible measures $Q_E$ (one class for each valid $E$ \cite{hororev}) in terms of the minimum entanglement $E$ necessarily created during measurement processes.

\smallskip
\noindent {\it Main result.}---
We are now ready to state the main result of this paper, namely the ordering between entanglement and general quantum correlations in the state of an arbitrary quantum system $S_\cI$.
\begin{theorem}
\label{thm:main}
All measures $Q_E$ of quantum correlations defined by Eq.~\eqref{defqeq} satisfy
\beq
\label{eq:main}
Q_{E}^{(S_I)}({\rho_{S_{\cI}}}) \geq Q_{E}^{(S_J)}({\rho_{S_{\cI}}})\geq E_{S_K:S_{\cI\backslash K}}({\rho_{S_\cI}}),
\eeq
for all corresponding entanglement measures $E$, and all choices of $K \subseteq J\subseteq I\subseteq\cI$. In particular, in the bipartite case, one has
\beq
\label{eq:mainbip}
\begin{split}
Q_E^{(AB)}({\rho_{AB}})&\geq \max\{Q_E^{(A)}({\rho_{AB}}),\,Q_E^{(B)}({\rho_{AB}})\}\\
&\geq \min\{Q_E^{(A)}({\rho_{AB}}),\,Q_E^{(B)}({\rho_{AB}})\}\\
&\geq E_{A:B}({\rho_{AB}}).
\end{split}
\eeq
\end{theorem}
A proof of Theorem~\ref{thm:main} is provided in the Appendix. The theorem is valid for all choices of $E$ as it relies only on the basic \emph{monotonicity} property
\beq
\label{eq:monotonicity}
E(\rho)\geq E(\Lambda_{\textrm{LOCC}}[\rho]),
\eeq
satisfied by any entanglement measure by definition \cite{hororev},
for all states $\rho$ and all transformations $\Lambda_{\textrm{LOCC}}$ that can be realized by Local Operations and Classical Communication (LOCC).

The leftmost inequality in \eqref{eq:main} tells us that the more subsystems are measured, the more quantum correlations are revealed in the state; this generalizes to arbitrary measures $Q_E$ the few known dominance relations involving, e.g., two-way versus one-way discord \cite{cavesrev}. The rightmost inequality tells us, crucially, that according to the framework of Definition \ref{defQ}, $Q_E$ and $E$ always obey a precise ordering relation, with the latter (however quantified), if present, accounting only for a fraction of the more general quantum correlations (compatibly defined) in arbitrary quantum states.
In particular, it is straightforward to check that, thanks to the Schmidt decomposition, the inequalities of \eqref{eq:mainbip} are saturated for pure states of the system, returning that entanglement and quantum correlations in general coalesce into a unique signature in absence of global mixedness \cite{discord}.
For generally mixed states, Eq.~\eqref{eq:main} puts the present approach to investigating non-classicality of correlations on firm physical grounds: {\em quantum correlations truly go beyond entanglement}, in a clear quantitative sense.


The rightmost sides of Eqs.~\eqref{eq:main} and \eqref{eq:mainbip} involve bipartite entanglement $E$, but we can also extend the ordering we have established to the multipartite case. For example, one can define two quite natural measures of (global) multipartite entanglement starting from any bipartite entanglement monotone $E$:
$E_{\textrm{min(max)}}(\rho_{S_\cI}):=\min(\max)_{\{P_0,P_1\}}E_{S_{P_0}:S_{P_1}}({\rho_{S_\cI}})$,
where the minimization (maximization) is taken over all non-trivial partitions of $\cI$, that is, $P_0\cap P_1 = \emptyset, P_0\cup P_1=\cI$. The two multipartite measures inherit the LOCC monotonicity \eqref{eq:monotonicity} from the bipartite measure $E$, and both coincide with the latter in the bipartite case. Then, from \eqref{eq:main} it follows for instance that $Q_E^{(S_\cI)}({\rho_{S_\cI}})\geq E_{\textrm{max}}(\rho_{S_\cI})\geq E_{\textrm{min}}(\rho_{S_\cI})$.


\smallskip
{\noindent \it Flow of quantum correlations and the von Neumann chain.}---
We now move to characterizing the spreading of entanglement among systems and measurement apparatuses. We will refer to the following notion of {\it genuine} multipartite entanglement.
\begin{definition}
A pure state $\ket{\psi}_{S_\cI}$ is genuinely multipartite entangled if $\ket{\psi}_{S_\cI}\neq \ket{\psi_0}_{S_{P_0}}\ket{\psi_1}_{S_{P_1}}$ for all non-trivial bipartitions $\{P_0,P_1\}$ of $\cI$. A mixed state $\rho_{S_\cI}$ is genuinely multipartite entangled if for any pure-state ensemble decomposition $\{p_i,\ket{\psi_i}_{S_\cI}\}$ (such that $\rho_{S_\cI} = \sum_i p_i \proj{\psi_i}_{S_\cI}$) there is at least one genuine multipartite entangled state appearing with non-vanishing probability.
\end{definition}
\noindent We get the following result, whose proof is in Appendix.
\begin{theorem}\label{thm:multi}
The state $\rho_{S_\cI}$ is genuinely multipartite entangled if and only if any pre-measurement state $\tilde\rho_{S_\cI M_I}$, for any $I\subseteq \cI$, is genuinely multipartite entangled, with multipartite entanglement shared among all subsystems of $S_\cI \cup M_I$.
\end{theorem}

All the above results can be used to analyze qualitatively and quantitatively a von Neumann chain whose links are given by a sequence of measurement apparatuses (see Fig.~\ref{fig:vNchain}). That is, suppose that a measurement apparatus $M_1$, initially in a pure state, is used to probe a single system $S$ in a state $\rho_S$. The measurement is performed by letting $S$ and $M_1$ interact as in Eq.~(\ref{eq:uniinteraction}), leading to a pre-measurement state $\tilde{\rho}_{SM_1}$. Using the same tools of~\cite{bruss,ouracti}, it is easy to check that $\tilde{\rho}_{SM_1}$ is $S:M_1$ entangled if and only if the measurement is not performed in the eigenbasis of $\rho_S$. We can now consider another link in the von Neumann chain, given by another apparatus $M_2$ brought in to realize a complete projective measurement on $M_1$. We thus obtain a global pre-measurement state $\tilde{\rho}'_{SM_1M_2}$. Theorem \ref{thm:main} and Definition \ref{defQ} imply $Q_E^{(M_2)}({\tilde{\rho}'_{SM_1M_2}}) \geq E_{SM_1:M_2}({\tilde{\rho}'_{SM_1M_2}}) \geq Q_E^{(M_1)}(\tilde{\rho}_{SM_1}) \geq E(S:M_1)_{\tilde{\rho}_{SM_1}}$. This argument can be reiterated for the next links in the chain, so that we find $E_{S:M_1}\leq E_{SM_1:M_2}\leq\ldots \leq E_{SM_1M_2\ldots M_n:M_{n+1}}$, where bipartite entanglement is calculated for larger and larger systems (that is, longer and longer chains, last link versus the rest). Similarly $Q_E^{(M_1)} \le \ldots \le Q_E^{(M_n)}$  for the quantum correlations of the corresponding successive pre-measurement states of the chain. We thus conclude that bipartite {\em entanglement and quantum correlations never decrease along von Neumann chains.}
Notice that it follows also that, when we consider the complete chain and break it at any level, namely investigating the bipartition $(S M_1 \ldots M_j):(M_{j+1} \ldots M_n)$, the entanglement and the corresponding quantum correlations across this (global) bipartition are both nondecreasing functions of $j$. Indeed, 
successive measurement steps correspond to a local operation (actually, a local isometry) with respect  to the fixed bipartition at the level $j$. 

These considerations are readily generalized to the case of $S$ being a composite system. Let us focus for simplicity on the bipartite case, with $B$ being measured by $M_1$, which is in turn being measured by $M_2$ and so on. We have then
\begin{eqnarray*}
E_{A:B}(\rho_{AB}) &\le& Q^{(B)}_E(\rho_{AB}) \le E_{AB:M_1}(\tilde\rho_{ABM_1}) \\
&\le& Q^{(M_1)}_E(\tilde\rho_{ABM_1})  \le E_{ABM_1:M_2}(\tilde\rho'_{ABM_1M_2}) \\
&\le& Q^{(M_2)}_E(\tilde\rho'_{ABM_1M_2}) \le \ldots\,.
 \end{eqnarray*}
 The first step of the chain is the most important one, where the nature of correlations in the initial state of $\rho_{AB}$ of the system plays a crucial role.  If such a state is $B$-CC  (i.e., $Q_E^{(B)}(\rho_{AB})=E_{A:B}(\rho_{AB})=0$), then there exists a local von Neumann measurement on $B$ such that the pre-measurement state $\tilde{\rho}_{ABM_1}$ contains no entanglement and no quantum correlations between the system and the involved apparatus \cite{bruss}. If the initial state of the system is unentangled but quantumly correlated \cite{footnotess} ($Q_E^{(B)}(\rho_{AB})>0, E_{A:B}(\rho_{AB})=0$), then during the pre-measurement the initial intra-system quantum correlations $Q_E^{(B)}(\rho_{AB})$ are transformed (and possibly amplified) into quantum correlations $Q_E^{(M_1)}(\rho_{ABM_1}) \ge Q_E^{(B)}(\rho_{AB})$ between the composite system and the apparatus, and correspondingly an amount of entanglement $E_{AB:M_1}(\rho_{ABM_1}) \ge Q_E^{(B)}(\rho_{AB})$ is also created across the $AB:M_1$ bipartition. 
Finally, according to Theorem \ref{thm:multi}, iff the initial state of the system is entangled ($Q_E^{(B)}(\rho_{AB})\geq E_{A:B}(\rho_{AB})>0$), then genuine multipartite entanglement is established between the two subsystems and the apparatus, and keeps being genuinely shared with all the subsequent apparatuses as well. 
Notice that after the initial step, the amount of bipartite entanglement between the latest apparatus and the rest of the chain can actually stay constant ($E_{AB:M_1}=\ldots = E_{ABM_1M_2\ldots M_n:M_{n+1}}$) if every apparatus realizes an optimal---in the sense of Eq. \eqref{defqeq}---local measurement.  This is seen by noticing that the $AB:M_1$ entanglement in $\tilde{\rho}_{ABM_1}=\sum_{i,j}\unidim{i}{j}_{A}\ot \rho^{ij}_{B}\ot\unidim{i}{j}_{M_1}$ is trivially the same as the $ABM_1:M_2$ entanglement in $\tilde{\rho}'_{ABM_1M_2}=\sum_{i,j}\unidim{i}{j}_{A}\ot \rho^{ij}_{B}\ot\unidim{i}{j}_{M_1}\ot\unidim{i}{j}_{M_2}$, the latter state being obtained by letting $M_1$ and $M_2$ interact as in Eq.~\eqref{eq:uniinteraction}.

A more realistic model of the measurement process would involve the system $\bf S$, a number of apparatuses ${\bf M}$, and a large set of uncontrollable degrees of freedom---the environment ${\bf E}$---that act by randomly measuring the subsystems and/or the apparatuses \cite{zurek91}. While we can have a fine control on the measurement apparatuses, tuning them to realize minimally-disturbing von Neumann measurements on the objects to be observed, the non-tailored action of the environment results in it getting strongly quantumly correlated with the accessible degrees of freedom through bipartite and genuine multipartite entanglement (subject to stringent ``monogamy'' constraints \cite{monogamy}), according to the mechanisms elucidated above.
After tracing over the unaccessible environmental degrees of freedom, the observed systems and the measurement apparatuses are thus typically left in classically correlated ``pointer states''. This is the essence of decoherence by environmental-induced selection \cite{zurekrmp}, and our findings allow us in principle to get a quantitative grip on the nature and measure of (quantum) correlations during the process. A detailed study of this mechanism requires further investigation, possibly considering also the case of ``fuzzy measurements'' in which the measurement apparatuses are themselves noisy, i.e., initialized in mixed rather than pure states \cite{vedralfuz}: in that case, it will be interesting to analyze the role of {\it classical} correlations as well as their interplay with entanglement and general quantum ones.

\smallskip
{\noindent \it Conclusions.}---
The results presented in this Letter provide physical justification and demonstrate the insightfulness of the operational approach put forward in \cite{bruss,ouracti} to address quantitatively the general quantum correlations in a composite system in terms of the entanglement necessarily created between the system and measurement apparatuses during local measurements. In this Letter, we proved the key result that, within such a framework, quantum correlations are never smaller than entanglement, going thus beyond it in general. This holds for any entanglement monotone and any correspondingly defined measure of quantum correlations. We provided quantitative and qualitative results on the presence and spreading of entanglement and quantum correlations along von Neumann chains. We hope that, motivated by our findings, coordinated efforts will lead soon to a comprehensive mathematical resource theory of general quantum correlations and to novel demonstrations of their power for noise-resilient quantum technologies \cite{merali}.

\smallskip
{\noindent \it Acknowledgements.}---
We thank V. P. Belavkin, D. Bruss, E. Chitambar, N. Gisin, T. Nakano, D. Terno and particularly J. Calsamiglia for discussions. MP acknowledges support by NSERC, CIFAR, Ontario Centres of Excellence. GA is supported by a Nottingham Early Career Research and Knowledge Transfer Award. We acknowledge joint support by the EPSRC Research Development Fund (Pump Priming grant 0312/09).

\appendix*

\section{Appendix: Proofs}

\subsection*{Proof of Theorem \ref{thm:main}}
For the sake of clarity, let us consider the bipartite case, with a local measurement applied only to $A$ through an apparatus $M$. The initial bipartite state of the system can be expanded as
$$\rho_{AB}=\sum_{i,j}\unidim{i}{j}_{A}\ot \rho^{ij}_{B}\,,$$ where $\{\ket{i}\}$ is the basis in which $A$ is measured. The pre-measurement state is then $$\tilde{\rho}_{AMB}=\sum_{i,j}\unidim{i}{j}_{A}\ot\unidim{i}{j}_{M}\ot \rho^{ij}_{B}\,.$$ The key observation is that there is a simple LOCC transformation $\Lambda_{\textrm{LOCC}}$ (with respect to the the $AB:M$ bipartite cut) such that $\Lambda_{\textrm{LOCC}}[\tilde{\rho}_{AMB}]=\rho_{MB}$, where  $\rho_{MB}$ is the same state as $\rho_{AB}$ but now shared between $M$ and $B$. Due to \eqref{eq:monotonicity} we have $$E(AB:M)_{\tilde{\rho}_{AMB}}\geq E_{M:B}({\rho_{MB}})=E_{A:B}({\rho_{AB}})\,.$$ It remains to exhibit the operation $\Lambda_\textrm{LOCC}$ on $\tilde{\rho}_{AMB}$, which can be constructed as follows. A Fourier transform $\ket{i}\mapsto1/\sqrt{d}\sum_k e^{2\pi {\rm i} ik /d}\ket{k}$, with $d$ the dimension of $A$ (and of $M$) is first applied to $A$, transforming the pre-measurement state into $$\frac1d\sum_{k,l}\sum_{i,j}e^{2\pi\ii(ik-jl)/d}\unidim{k}{l}_{A}\ot\unidim{i}{j}_{M}\ot \rho^{ij}_{B}\,.$$ Then subsystem $A$ is measured in the $\{\ket{k}\}$ basis, and depending on the result a unitary $U_k=\sum_ie^{-2\pi\ii ik/d}\proj{i}$ is applied to $M$, obtaining $$\tilde{\rho}_{MB}=\sum_{i,j}\unidim{i}{j}_{M}\ot \rho^{ij}_{B}\,.$$

The proof can be straightforwardly generalized to the multipartite case.
Given a pre-measurement state ${\tilde{\rho}_{S_\cI M_I}}$
there are $S_k:M_k$ LOCC transformations that ``undo'' the measurement interaction between $S_k$ and $M_k$, for any $k\in I$. Indeed, it is possible to act via LOCC to  choose between the following two options: (i) to really just undo the measurement interaction on $S_k$ (as if no apparatus $M_k$ had been introduced at all) or (ii) to transfer coherently the quantum information contained originally in $S_k$ into $M_k$. Option (i) justifies the first inequality in Eq.~\eqref{eq:main}; option (ii) justifies the second one, as for any system $S_k$ interacting with its measuring apparatus (i.e., $k\in I$, or $k\in J$ in Eq.~\eqref{eq:main}) we can choose to have its information content transferred to $M_k$, on the other side of the $S_\cI : M_I$ splitting. The latter possibility means that we can go from the the pre-measurement state $\tilde{\rho}_{S_\cI:M_I}$ to $\rho_{S_{\cI\backslash K}: M_{K}}$, for all $K\subseteq I$. The latter state is the same as the original $\rho_{S_\cI}$, only with the information contained in the systems $S_K$ transferred to the apparatuses $M_K$.
\hfill $\square$

\subsection*{Proof of Theorem \ref{thm:multi}}
To prove the theorem we can restrict to measurements on just one subsystem, say $S_n$, since we are considering multipartite entanglement between all subsystems and apparatuses, and we can analyze the measurement on each $S_k$ separately.
Following \cite{john}, we observe that pure-state decompositions for $\rho_{S_\cI}$ and $\tilde{\rho}_{S_\cI M_n}$ are in one-to-one correspondence, as the initial state of $M_n$ is pure (for each $\ket{\psi}_{S_\cI}$ we obtain a pure state $\ket{\tilde{\psi}}_{S_\cI M_n}=U_{S_n M_n}\ket{\psi}_{S_\cI}\ket{0}_{M_n}$, and for each $\ket{\tilde{\psi}}_{S_\cI M_n}$ there must be a state $\ket{\psi}_{S_\cI}$ such that $U^\dagger_{S_n M_n}\ket{\tilde{\psi}}_{S_\cI M_n}=\ket{\psi}_{S_\cI}\ket{0}_{M_n}$).
One direction is then trivial: if $\rho_{S_\cI}$ is not genuine multipartite entangled, $\tilde{\rho}_{S_\cI M_n}$ is not genuine multipartite entangled either. Indeed, the measurement interaction will map a state $\ket{\psi_0}_{S_{P_0}}\ket{\psi_1}_{S_{P_1}}$, for $\{P_0,P_1\}$ a partition of $\cI$, into a state $\ket{\psi_0}_{S_{P_0}}\ket{\tilde{\psi}_1}_{S_{\tilde{P}_1}}$, with $\{P_0,\tilde{P}_1\}$ a new partition including $M_n$ in $S_{\tilde{P}_1}$, if we assume, without loss of generality, that $S_n\in P_1$. We now prove also that if $\tilde{\rho}_{S_\cI M_n}$ is not genuine multipartite entangled, then $\rho_{S_\cI}$ was not either.

Let us consider a pure state $\ket{\tilde{\psi}}_{S_\cI
M_n}=\ket{\tilde{\psi}_0}_{S_{\tilde{P}_0}}\ket{\tilde{\psi}_1}_{S_{\tilde{P}_1}}$ in the decomposition of $\tilde{\rho}_{S_\cI M_n}$. If both $S_n$ and $M_n$ are in $\tilde{P}_1$, then undoing the measurement interaction we obtain a state $\ket{{\psi}_0}_{S_{{P}_0}}\ket{{\psi}_1}_{S_{{P}_1}}$. Suppose instead $S_n \in \tilde{P}_0$, $M_n \in \tilde{P}_1$. Imposing $$\ket{\tilde{\psi}}_{S_\cI M_n}=\ket{\tilde{\psi}_0}_{S_{\tilde{P}_0}}\ket{\tilde{\psi}_1}_{S_{\tilde{P}_1}}=U_{S_n M_n}\ket{\psi}_{S_\cI}\ket{0}_{M_n}\,,$$ namely imposing in particular that $\Tr_{S_{\tilde{P}_{1}}}(\proj{\tilde{\psi}}_{S_\cI M_n})$ is pure, we find that $S_n$ must be initially unentangled from the other subsystems and actually in an eigenstate of the measurement performed by $U_{S_n M_n}$.
 \hfill $\square$


\begin{thebibliography}{99}
\bibitem{footnotess}
For the sake of the discussion we  consider here a faithful entanglement measure $E$ such that $E(\rho_{AB})=0$ iff $\rho_{AB}$ is not entangled.
\end{thebibliography}
\end{document}